\documentclass[envcountsect, runningheads]{llncs}
\usepackage{graphicx}
\usepackage{cite}
\usepackage{listings}
\usepackage{amsmath}
\usepackage{mathtools}
\usepackage{ebnf}
\usepackage{lmodern}
\usepackage{todonotes}
 
 \usepackage[frozencache,cachedir=.]{minted}
 \usepackage{placeins}
 \usepackage{wrapfig}
\usepackage[citecolor=blue, urlcolor=blue, colorlinks=true, linkcolor=blue]{hyperref}

\begin{document}
\title{A formal approach for customization of schema.org based on SHACL}
\titlerunning{A formal approach for customization of schema.org}
%
\author{Umutcan \c{S}im\c{s}ek\orcidID{0000-0001-6459-474X} \and
Kevin Angele\orcidID{1111-2222-3333-4444} \and
Elias K\"{a}rle\orcidID{0000-0002-2686-3221} \and Oleksandra Panasiuk \and Dieter Fensel}
\authorrunning{U. Simsek et al.}
%
\institute{
University of Innsbruck, Innsbruck, Austria \\
\email{\{firstname.lastname\}@sti2.at}\\}
\maketitle              
%
\begin{abstract}
Schema.org is a widely adopted vocabulary for semantic annotation of content and data. However, its generic nature makes it complicated for data publishers to pick right types and properties for a specific domain and task. In this paper we propose a formal approach, a domain specification process that generates domain specific patterns by applying operators implemented in SHACL to the schema.org vocabulary. These patterns can support knowledge generation and assessment processes for specific domains and tasks. We demonstrated our approach with use cases in tourism domain. 
\end{abstract}
%
%
%
\section{Introduction}

Semantic annotation of content and data is a fundamental task for generating the building blocks of many endeavors on the semantic web such as building knowledge graphs, generally speaking, making the content consumable by automated agents.

\begin{figure}[h]
	\centering
	\includegraphics[width=0.8\textwidth]{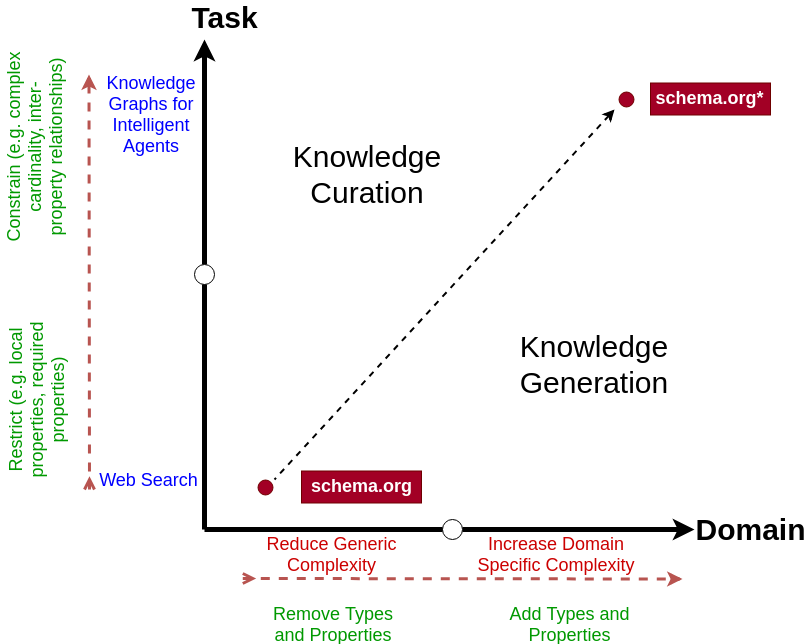}
	\caption{schema.org vocabulary in task and domain dimensions. The goal of domain specification process is to create patterns that are domain and task specific.}
	\label{fig:ontology-dimensions}
\end{figure}

 Schema.org vocabulary \cite{journals/cacm/GuhaBM16} is currently the de-facto standard for annotating content and data on the web. The vocabulary is maintained by the schema.org initiative and it contains 614 types and 905 properties\footnote{See \url{https://schema.org/docs/schemas.html} Last accessed: 02.04.2019}. In terms of coverage, schema.org can be examined in two dimensions, namely task and domain. Figure \ref{fig:ontology-dimensions} demonstrates the schema.org vocabulary between these dimensions. The x-axis represents the domain dimension, ranging from high generic complexity to high domain-specific complexity. High generic complexity indicates that a vocabulary covers various domains in a superficial way. As the domain-specific complexity increases, the number of domains covered by a vocabulary decreases and the level of detail of a specific domain increases. The y-axis represents the task dimension, ranging from web search to building knowledge graphs. These tasks contain various processes such as knowledge generation and curation. Knowledge generation is creating semantically annotated data based on (a) manual (e.g. via a web form) (b) semi-automated (e.g. via mappings on metadata of structured sources) and (b) automated (e.g. via extraction from unstructured data) techniques. Knowledge curation is a process consists of knowledge assessment, knowledge cleaning and knowledge enrichment \cite{DBLP:conf/sigmod/DongS15, DBLP:journals/semweb/Paulheim17}.  On the task dimension, schema.org vocabulary can be found closer to the web search task, since it contains types and classes to represent everything relevant for this task. On the domain dimension, it has high generic complexity, since it contains types and properties to represent several domains (e.g. events, media, accommodation) in a shallow manner.

 The schema.org vocabulary aims to achieve rapid uptake and growth, and to that end, they take the burden off of the content publisher and put it on the data consumer. Therefore its data model is quite flexible in terms of type hierarchy and inheritance of properties by specialized types (e.g. a Waterfall can have a telephone number). Moreover, there are multiple ways to represent the same information. This is achieved by using disjunctive ranges  (e.g. the address of a local business can be defined as a string value or as a structured value) and allowing multiple properties with the same purpose (e.g. address of local business can be defined through address, contactPoint or location property) on a type. Although this flexibility and heterogeneity serve for the goals of schema.org vocabulary, it comes with a side effect of contributing to generic complexity of the vocabulary. The high generic complexity can make the knowledge generation process quite challenging, especially within a specific domain. This can happen for two reasons (a) the publishers may not know which types and properties are correct for their domain (b) the schema.org vocabulary may not contain necessary types and properties for their domain. From a task perspective, the vocabulary may need to be restricted and constrained in a certain way that it fits to the needs of a specific task. For instance, a domain specific pattern can be used to support knowledge curation, to assess whether the knowledge graph contains the required information in a way that fits to the requirements of an application like a dialogue system.
 
 One way to guide data publishers to create semantically annotated data and content for a specific domain and task is to provide domain specific patterns of the schema.org vocabulary. On one hand these patterns can make schema.org vocabulary suitable for different tasks by restricting and constraining the vocabulary, on the other hand, they can make it more domain specific by reducing the generic complexity and increasing the domain specific complexity. These patterns would serve as an agreement between publishers and consumers to ease the knowledge generation and curation processes for specific domains and tasks. In this paper we present a formal approach for domain specification, a process that applies an operator to the schema.org vocabulary in order to generate a domain specific pattern. We give formal definitions of domain specification operators and an abstract syntax based on Shapes Constraint Language (SHACL) \cite{bibSHACL}. In order to demonstrate its usage, we give concrete domain specification operator implementations in SHACL and introduce several use cases.

The rest of the paper is structured as follows: Section \ref{sec:schema-org} gives a brief introduction to the schema.org and describe the formalization we adopt. In Section \ref{sec:shacl}, we give a brief introduction to SHACL and its components we adopt for defining domain specification operators. Section \ref{sec:domain-spec} describes the domain specification process formally. We show some concrete domain specification examples for tourism domain in Section \ref{sec:examples}. Section \ref{sec:related-work} gives an overview of the related work and Section \ref{sec:conclusion} concludes the paper with final remarks and future directions.

\section{Schema.org}
\label{sec:schema-org}
schema.org is a vocabulary to describe "things" on the web. With this intention it was founded in 2011 by four major search engine providers Bing, Google, Yahoo! and Yandex. The development of schema.org is not only driven by a consortium that consists of members of all four mentioned companies but also by a large community, that collaborates on the vocabulary. The vocabulary has regular releases\footnote{v3.5 as of 02.04.2019}. Each release addresses some issues about the existing types and properties or extends the vocabulary by adding new types and properties. Besides extending the core vocabulary directly, third parties can define "external extensions" under a different namespace. These extensions typically contain types and properties that are aligned with the core schema.org vocabulary\footnote{e.g. with RDFS or SKOS constructs }.   

The data model of schema.org is quite simple and, according to the schema.org documentation, "very generic and derived from RDF Schema"\footnote{https://schema.org/docs/datamodel.html - accessed on April 1, 2019}. The schema.org data model organizes types in a multiple inheritance hierarchy. The vocabulary contains properties that have one or more types in their domains and one or more types in their range definitions. The ranges of a property may need to be altered in practice, by defining a value with multiple types. A prominent example for this so called multi-typed entity (MTE)\footnote{\url{https://github.com/schemaorg/schemaorg/wiki/How-to-use-Multi-Typed-Entities-or-MTEs}} practice is the annotation of hotel rooms \footnote{\url{https://schema.org/docs/hotels.html}}. The schema:HotelRoom type contains properties for describing beds and amenity features. However, to describe a daily room price, the hotel room must be defined also as an instance of schema:Product, which allows the usage of schema:offers property. The MTE practice is a good example for why customization of schema.org for specific domains and tasks is needed.  The MTE practice is in principle a way to overcome the generic nature of schema.org vocabulary for specific domains and tasks. However,  for data publishers in tourism domain, it may be complicated to find out which types should be used for an MTE. Domain specific patterns created by domain experts can enforce conjunction of multiple types as the range of a property to guide data publishers.

As for the formal definition of schema.org, we mostly share the views presented by Patel-Schneider in \cite{DBLP:conf/semweb/Patel-Schneider14}. Following that full-fledged definition, we make the below formal definition of schema.org for the purpose of describing the domain specification process described in this paper:

\begin{definition} The schema.org vocabulary is a pair $(T,P)$ of finite sets $T$ and $P$, where:  
	
\begin{enumerate}
\item $T \cap P = \emptyset$.
\item a type $t \in T$ is a regular type when $t$ is a more specific type than Thing in schema.org vocabulary
\item a type $t \in T$ is a datatype when $t$ is a more specific type than DataType in schema.org vocabulary.
\item each type and property is identified with a URI in the schema.org namespace
\item  a type in $T$ or a property in $P$ can be represented with only one URI and a URI that represents a type or a property cannot represent another type or a property.
\end{enumerate}
\end{definition}

\begin{definition} A type $t \in T$ is a pair $(U, S)$  where $U$ is the identifier of the $t$, $S$ is the set of subtypes of $t$.
\end{definition}
\begin{definition} A property $p \in P$ is a pair $(U, D, R)$  where U is the identifier of $p$ and D and R are the set of types that are in the domain and range of $p$, respectively.
\end{definition}






While the definitions above are inspired by the formal definition made in  \cite{DBLP:conf/semweb/Patel-Schneider14}, we introduce some simplifications based on the scope of our work. The work in \cite{DBLP:conf/semweb/Patel-Schneider14} defines a schema.org knowledge base, which consists of types, properties and instances. Since we only deal with the vocabulary for creating domain specific patterns, we ignore the instances. We consider enumeration types, as regular types in $T$. The distinction between datatypes and non-datatypes are only made for syntactic purposes to support SHACL syntax (See Section \ref{sec:domain-spec}). Additionally, we do not consider subproperties in our definition. 



\section{SHACL}
\label{sec:shacl}

SHACL is a recent W3C recommendation for defining constraints over RDF data. The language does not have a normative abstract syntax and formal semantics\footnote{See \href{https://www.w3.org/TR/shacl-abstract-syntax/}{here} for details and a link to a proposed abstract syntax and semantics. Last accessed on 23.03.2019}. A syntactically well-formed SHACL shape is ensured by a set of shapes\footnote{https://www.w3.org/ns/shacl-shacl} that implement the SHACL Syntax Rules \footnote{https://www.w3.org/TR/shacl/\#syntax-rules}. For the domain specification process proposed in this paper, we use a subset of SHACL-CORE\footnote{\textit{sh} prefix is used for SHACL-Core namespace} elements. 
We adopt the following SHACL elements for the syntax of domain specification operators:  

\begin{enumerate}
	\item Node shapes (sh:NodeShape) to define restricted types
	\item Property shapes (sh:PropertyShape) to define allowed properties on a restricted type
	\item Class-based targets (sh:targetClass) to determine the target type of a domain specification operator 
	\item Value type constraints (sh:class and sh:datatype) to modify the range of an allowed property on a restricted type.
	\item Cardinality constraints to define required properties
	\item Shape-based (sh:node and sh:property) constraints to define range restrictions
	\item Logical constraints (sh:or) to apply logical operations on a set of constraints
\end{enumerate}

The aforementioned elements are selected based on our experience with schema.org annotations in various use cases in tourism domain (see Section \ref{sec:examples}). We explain how these elements are used in our abstract syntax for domain specification operators in Section \ref{sec:domain-spec}. The elements we adopt from SHACL vocabulary and their dependencies (e.g. URI and blank node syntax as defined in RDF specification) do not imply any semantics for our approach. 

\section{Domain Specification} 
\label{sec:domain-spec}


Domain specification is a process that generates domain specific patterns by applying an operator to the schema.org vocabulary. The resulting patterns provide basis for knowledge generation from structured and unstructured sources and help to assess knowledge graphs on a larger scale.

Overall, domain specification process has two major goals in the domain and task dimensions (see Figure \ref{fig:ontology-dimensions}) (a) reducing the generic complexity and increasing domain specific complexity (b) restricting and constraining the schema.org vocabulary according to certain tasks. We define following types of domain specification processes to achieve these goals:
\begin{itemize}
    \item Simple Domain Specification (SDS) reduces the generic complexity of schema.org by generating a Simple Domain Specific Pattern (SDSP) through removing types and properties.
    \item Restriction Domain Specification (RDS) generates a Restricted Domain Specific Pattern (RDSP)  by restricting the ranges of remaining properties in an SDSP for various tasks
    \item Extension Domain Specification (EDS) generates an Extended Domain Specific Pattern (EDSP) that increases the domain-specific complexity by adding new types and properties to an SDSP or RDSP
\end{itemize}

In the rest of this section, we give formal definitions for the three of four aforementioned domain specification process types, namely SDS, RDS and EDS. We explain the patterns produced by these processes with a running example and describe the abstract syntax of domain specification operator for each process based on SHACL elements. \footnote{We describe the abstract syntax for domain specifications with \href{https://www.w3.org/TR/2004/REC-xml-20040204/\#sec-notation}{"EBNF for XML" notation}. We extend the notation with [..] structure for describing special sequences in natural language (e.g. [a valid URI]).} The definition of CDS is left for the future work. 

\subsection{Simple Domain Specification (SDS)}
\label{sec:sds}
An SDS produces an SDSP based on schema.org through following steps:
\begin{enumerate}
    \item Define $T^{\cap}$, a subset of $T$ by excluding types and $P^{\cap}$, a subset of $P$ by excluding properties from the schema.org vocabulary (Equation \ref{eq:subsets}). 
    \begin{equation}
    \label{eq:subsets}
    \begin{gathered}
        T^{\cap} \subseteq T \\
        P^{\cap} \subseteq P  
    \end{gathered}
    \end{equation}
	
	\item Define $T_p$, the local properties for types in $T^{\cap}$ (Equation \ref{eq:localprops}): 
	\begin{equation}
	\label{eq:localprops}
	\begin{gathered}
	T_p = \{(type, property) \in T^{\cap} \times P^{\cap} \colon domain^\cap(property,type) \}\\ 
	\end{gathered}
	\end{equation}
    
     The predicate "$domain^\cap$" holds true for a $type \in T^{\cap}$ and $p = (U,D,R)$ where $type \in D$.
     
     The following example shows the local properties for the type \texttt{Hotel}\footnote{s is a prefix for http://schema.org/ namespace.} in an SDSP.
     \begin{equation}
         \begin{gathered}
             (\{(s:Hotel, s:checkinTime), (s:Hotel, s:checkoutTime), \\(s:Hotel, s:containsPlace), (s:Hotel, s:location), \\(s:Hotel, s:name)\})
         \end{gathered}
     \end{equation}

    \item Define  $T_r$, the local ranges defined on a local property on a type (Equation \ref{eq:localranges}): 
     \begin{equation}
    \label{eq:localranges}
    \begin{gathered}
    T_r = \{((type, property), rangeType) \in T_p \times T^{\cap} \colon\\ range^\cap(property, rangeType) \}\\ 
    \end{gathered}
    \end{equation}
    
     The predicate "$range^\cap$" holds true for a $type \in T^{\cap}$ and $p = (U,D,R)$ where $type \in R$.

 Simple local ranges are defined to build an SDSP as follows\footnote{ All subtypes of the shown types ($type = (U, S)$ and $type_{sub} \in S$) are included in all three types of domain specific patterns}:

     \begin{equation}
        \begin{gathered}
            (\{((s:Hotel, s:checkInTime), s:DateTime), \\((s:Hotel, s:checkOutTime), s:DateTime),\\ ((s:Hotel, s:containsPlace), s:Place), \\((s:Hotel, s:location), s:PostalAddress),\\((s:Hotel, s:name), s:Text)\})
        \end{gathered}
    \end{equation}
\end{enumerate}

 The process described and demonstrated above is realized through an operator defined in terms of SHACL elements. The abstract syntax of this operator is as follows:
 
An SDS operator is a node shape that is identified with a URI or a blank node. An SDS operator has a target type from the schema.org vocabulary and contains one or more property shapes, each of which represents the definition of a local property on the target schema.org type.

\begin{EBNF}
	
	\item[SDS]
	\<NodeShape>, \<SDOTargetType>, \<SDSPropertyShape>+;
	\item [NodeShape]
	\<Identifier>, \<NodeShapeType>;
	\item[Identifier] 
	\spec {a URI}  | \spec {a BlankNode};
	\item[NodeShapeType]
	\spec{ URI of the SHACL NodeShape Type }
	\item[SDOTargetType] 
	\spec{a class-based target whose value is a URI of type t is a regular type in T};
\end{EBNF}
A property shape in an SDS consists of a schema.org property and one or more type specifications for the range of the property. It may also optionally contain a cardinality constraint to specify whether a property is required. A property shape may specify a range with multiple disjunctive types.
\begin{EBNF}
	\item[SDSPropertyShape] 
	\<SDOProperty>, \<CardinalityConstraint>?, (\<ValueTypeConstraint+> | \<SimpleDisjunctiveConstraint>);
	\item[SDOProperty]
	\spec{URI of property p where $p \in P$~};
\end{EBNF}

A value type constraint enforces the type of a value. It requires the URI of a type in schema.org vocabulary as the value of sh:class parameter or the URI of an XSD datatype as the value of sh:datatype property.
\begin{EBNF}
	\item [ValueTypeConstraint]
	\<SimpleClassConstraint> | \<DatatypeConstraint> ;
	\item[SimpleClassConstraint]
	\spec{A SHACL Class constraint with sh:class parameter and a value t where t is a regular type in T};
	\item[DatatypeConstraint]
	\spec{A SHACL Datatype constraint with sh:datatype parameter and a value t where t is a datatype in T};
\end{EBNF}
A cardinality constraint enforces the number of values a property can take. It takes only the integer 1 as value for sh:minCount parameters to specify minimum required value occurrences.
\begin{EBNF}
	\item[CardinalityConstraint]
	\<MinCount>;
	\item[MinCount]
	\spec{minimum cardinality constraint with sh:minCount 1};
\end{EBNF}
A simple disjunctive constraint applies logical OR operation to a given list of value type constraints.
\begin{EBNF}
	\item[SimpleDisjunctiveConstraint]
	\spec{ OR(\<ValueTypeConstraint+>) } ;
	
\end{EBNF}
\subsection{Restricted Domain Specification (RDS)}
\label{sec:rds}
An RDS produces a restricted domain specific pattern by replacing one or more types in the range definitions of an SDSP with types from an external vocabulary. An external type (i.e. an element of the set $T_{ext}$) must be downwards compatible with the schema.org type that it replaces (i.e. the schema.org type could be used instead, in the data created based on the pattern). A type in $T_{ext}$ may be a part of an external extension of schema.org, meaning it is placed in schema.org's hierarchy. An external vocabulary is a pair $(T_{ext}, P_{ext}) $, where the types and properties are formalized in the same manner as schema.org vocabulary described in Section \ref{sec:schema-org}.

An RDS defines $T'_r$, by replacing one or more elements of $T_r$ with a \\ $((type, property), rangeType')$ where $(type,property) \in T_p$ is the type-property pair whose range will be restricted and $rangeType' \in T_{ext}$.

Assume there is a new type \texttt{n:HotelRoomProduct} \footnote{n is a prefix for an external vocabulary namespace}defined which is a conjunction of types \texttt{s:HotelRoom} and \texttt{s:Product} and is an element of $T_{ext}$. The type \texttt{n:Location} that is based on \texttt{s:Place} contains only the property \texttt{s:address}. Then, an RDF range restriction can be written as following\footnote{n:HotelRoomProduct is in the domain of all properties of s:HotelRoom and s:Product types. These properties, as well as all properties of s:PostalAddress are also in the domain specific pattern, but left out in the example for the sake of conciseness.}:

     \begin{equation}
\begin{gathered}
(\{((n:Location, s:address), s:PostalAddress),\\((s:Hotel, s:checkInTime), s:DateTime), \\((s:Hotel, s:checkOutTime), s:DateTime),\\ ((s:Hotel, s:containsPlace), n:HotelRoomProduct), \\((s:Hotel, s:location), n:Location),\\((s:Hotel, s:name), s:Text)\})
\end{gathered}
\end{equation}


The abstract syntax of an RDS operator is specified based on the abstract syntax of an SDS operator as follows:

An RDS is a SHACL node shape with a target type from the schema.org vocabulary and one more property shapes.

\begin{EBNF}
	\item[RDS]
	\<NodeShape>, \<SDOTargetType>, \<RDSPropertyShape>+;
\end{EBNF}

A property shape in an RDS extends a property shape in SDS with range restrictions.
\begin{EBNF}
	
	\item[RDSPropertyShape]
	\<SDOProperty>, \<CardinalityConstraint>?, (\<ValueTypeConstraint> | \<RangeConstraint> | \<DisjunctiveConstraint>)+;
\end{EBNF}
A range constraint consists of a value type constraint that declares a type in the range and a node constraint that further restricts the specified type.
\begin{EBNF}
	
	\item[RangeConstraint]
	\<ValueTypeConstraint>, \<NodeConstraint>;
	\item[NodeConstraint]
	\<NodeShape>, \<RDSPropertyShape>+;
\end{EBNF}
A disjunctive constraint extends a simple disjunctive constraint in an SDS by applying logical OR operator on a list of value type constraints and range restrictions.
\begin{EBNF}
	\item[DisjunctiveConstraint]
	\spec{OR\{(\<ValueTypeConstraint> | \<RangeConstraint>)+\}} ;
\end{EBNF}

\subsection{Extended Domain Specification (EDS)} 
An EDS process generates an extended domain specific pattern by extending an SDSP or an RDSP through
\begin{enumerate}
    \item adding new types and properties to define $T^{\cap\cup}$ and $ P^{\cap\cup}$ (Equation \ref{eq:eds-a})
    \item defining new properties to an existing type to define $T^{\cup}_p$ (Equation \ref{eq:eds-b})
    \item adding new types to the ranges of properties on an existing type to define $T^{\cup}_r$ (Equation \ref{eq:eds-c})
\end{enumerate}

The types and properties may come from an external vocabulary as described in Section \ref{sec:rds}.
There are two predicates in Equation \ref{eq:eds-b} and \ref{eq:eds-c} that help to build the sets for new properties for an existing type and new ranges on existing properties of existing types. The predicates $domain'$ and $range'$ hold true for the same conditions as defined for $domain^\cap$ and $range^\cap$ in Section \ref{sec:sds}, but for an external vocabulary instead of schema.org. Remember that, an external vocabulary can be an external extension of schema.org.

\begin{equation}
\label{eq:eds-a}
\begin{gathered}
    T^{\cap\cup} = T^{\cap} \cup \{type' \in T_{ext}\}\\
    P^{\cap\cup} = P^{\cap} \cup \{property' \in P_{ext}\}
\end{gathered}    
\end{equation}

\begin{equation}
\label{eq:eds-b}
\begin{gathered}
    T^{\cup}_p = T_p \cup \{(type', property') \in T_{ext} \times P_{ext} \colon domain'(property', type')\}
\end{gathered}    
\end{equation}

\begin{equation}
\label{eq:eds-c}
\begin{gathered}
    T^{\cup}_r = T_r \cup \{((type', property'), rangeType') \in T^{\cup}_p \times P_{ext} \colon range'(property', type')\}
\end{gathered}    
\end{equation}

The next example shows the definition of a property from an EDSP. It adds a new property, \texttt{n:totalNumberOfBeds}, to the type \texttt{s:Hotel} from an external extension of schema.org.  
     \begin{equation}
\begin{gathered}
(\{((n:Location, s:address), s:PostalAddress),\\((s:Hotel, s:checkInTime), s:DateTime), \\((s:Hotel, s:checkOutTime), s:DateTime),\\ ((s:Hotel, s:containsPlace), n:HotelRoomProduct), \\((s:Hotel, s:location), n:Location),\\((s:Hotel, s:name), s:Text)\\((s:Hotel, n:totalNumberOfBeds), s:Number))\})
\end{gathered}
\end{equation}




An EDS operator extends an SDS or an RDS operators with types from an external vocabulary. Given the abstract syntaxes of these two operators, an EDS can be described as follows:

\begin{EBNF}
	\item[EDS]
	\<NodeShape>, (\<SDOTargetType> | \<ExtTargetType>), \<EDSPropertyShape>+;
	\item[ExtTargetType]
	\spec{URI of type t where $t \in T_{ext}$};
	\item[EDSPropertyShape] 
	(\<SDOProperty> | \<ExtProperty>), \<CardinalityConstraint>?, (\<ExtValueTypeConstraint> | \<ExtRangeConstraint> | \<ExtDisjunctiveConstraint>)+;
	\item[ExtProperty]
	\spec{URI of property p where $p \in P_{ext}$};
	\item[ExtValueTypeConstraint]
	\<SimpleClassConstraint> | \<ExtClassConstraint> | \<DatatypeConstraint>;
	\item[ExtClassConstraint]
	[A SHACL Class constraint with a value t where $t \in T_{ext}$];
	\item [ExtRangeConstraint]
	\<ExtValueTypeConstraint>, \<ExtNodeConstraint>;
	\item[ExtNodeConstraint]
	\<NodeShape>, \<EDSPropertyShape>+;
	\item[ExtDisjunctiveConstraint]
	\spec{OR\{(\<ExtValueTypeConstraint> | \<ExtRangeConstraint>)+\}}
	
\end{EBNF}

\section{Use Cases and Examples}
\label{sec:examples}
In the section, we demonstrate domain specifications with examples and explain our current use cases and tooling. First we describe two different use cases where domain specific patterns have proven useful. Then we exemplify the domain specification process, particularly a concrete domain specification operator implemented with SHACL.
\subsection{Web Content Annotations in Tourism}
Domain specific patterns have been used in tourism domain by Destination Management Organizations (DMOs). The work in \cite{DBLP:conf/otm/AkbarKPSTF17} describes the generation of schema.org annotated tourism related data such as events, accommodation and infrastructure from raw data sources. The patterns have been used to guide the mapping process of the metadata of the IT solution provider of Mayrhofen DMO. The domain specifications are also used for manual generation of annotated data, through an editor that dynamically builds forms for annotation creation\footnote{https://actions.semantify.it/annotation/template/} based on domain specific patterns. The generated annotations have been used for the annotation of web pages for Search Engine Optimization and as a data source for a chatbot. The annotations created with the help of the domain specific patterns are evaluated qualitatively by observing search engine results. This use case shows a good example of making schema.org to fit a domain and suitable for different tasks such as consumption by chatbots.
\subsection{DACH-KG and Schema-Tourism Working Groups}
The DACH-KG working group\footnote{In German: https://www.tourismuszukunft.de/2018/11/dach-kg-auf-dem-weg-zum-touristischen-knowledge-graph/} was founded with the primary goal of building a touristic knowledge graph for the region of Austria, Germany and Switzerland (the DACH region). Several stakeholders in the tourism sector in the consortium is currently working on a unified schema to represent their data in the knowledge graph.  Hence the current focus of this working group lies on identifying the mappings from their data to schema.org and defining best practices for further use of schema.org. Domain specifications help the DACH-KG to formalize their findings and disseminate their best practice patterns to data providers in their field.

The schema-tourism working group was founded as a place for domain experts in tourism and researchers to commonly work on a) a unified way to use schema.org in the tourism domain and b) to identify shortcomings of schema.org and extend the vocabulary when needed. The findings of the working group are published as domain specific patterns and published online \footnote{https://ds.sti2.org}. There are currently 70 patterns created by the working group.  The documentation follows the schema.org documentation style and provides a list of all available domain specific patterns including a short description. By clicking the title of a pattern, its description page is shown. This page lists all mandatory and recommended properties with their respective ranges. If a type in the range is restricted in the pattern, then a link leads to the respective description. Otherwise, the link leads to schema.org's description of the type.
 
 \subsection{Domain Specification Example}
 In this section we present the domain specification operator created by schema-tourism working group implemented in SHACL. The operator generates a domain-specific pattern for hotel data based on schema.org\footnote{The full pattern can be found here: https://ds.sti2.org/Sypf3bVG1z/Hotel} (Listing \ref{lst:example-rds}). 
 Syntactically, the domain specification operator is a well-formed SHACL shape, therefore it can also be used as an input to a SHACL validator implementation. The datatypes of schema.org vocabulary are mapped to XSD datatypes in order to ensure the compatibility with existing tooling for SHACL. The example is an implementation of an RDS operator, however it is trivial to convert it to an EDS operator with an addition of a type or a property from an external vocabulary. 
 
 The example operator in Listing \ref{lst:example-rds} defines five local properties on type Hotel and removes all other properties from the type. All defined properties are required except containsPlace. Then, the local ranges are defined on those locally defined properties in terms of value-type and shape-based constraints. The operator further restricts the ranges location and containsPlace properties by replacing the types in the original ranges. First, it replaces the range of the containsPlace property Place, with a type that is the conjunction of both Product and HotelRoom. Second, it further restricts the range of the location property by replacing it with a type that is based on Place that only allows the address property. The range of the address property is also restricted to allow only the properties addressCountry and addressLocality. The example here demonstrates the motivation for using local ranges. For a given task (e.g. a dialogue system for booking hotel rooms), a hotel contains only hotel rooms. Therefore it makes sense to restrict the range of the containsPlace property on a Hotel to the HotelRoom type. Since hotel rooms can also have offers, the conjunction of HotelRoom and Product types are needed for the range. This allows the usage of all properties of Product, including the offers property. On the domain dimension, generic complexity is reduced by eliminating different ways of representing an address of a Hotel and allowing it only via the location property.

 \begin{listing}[h]
 	
 	\begin{inputminted}
 		[
 		fontsize=\scriptsize,
 		bgcolor=white,
 		breaklines,
 		frame=single,
 		]
 		{turtle}
 		{examples/rds.ttl}
 	\end{inputminted}
 	\caption{An example domain specification operator based on schema:Hotel type for tourism domain}
 	\vspace{-4mm}
 	\label{lst:example-rds}
 \end{listing}

\FloatBarrier
\section{Related Work}
\label{sec:related-work}
 Although there are not many tools and approaches targeting specifically restricting schema.org for content annotation, an informal definition of domain specification is made in \cite{10.1007/978-3-319-74313-4_31}. This approach only takes a subset of schema.org by removing types and properties. We define the domain specification process formally, and make clear conceptual distinction between the process of domain specification and resulting domain-specific patterns. Additionally, we do not only restrict but also allow extension of the schema.org vocabulary. 
 In a similar direction a plethora of approaches for validating RDF data have been proposed. A popular direction that has been explored by researchers is defining constraints with SPARQL. The work in \cite{conf/bis/FurberH10} utilizes SPARQL for defining domain independent constraints to check various data quality dimensions. RDFUnit is a framework that mimics the unit tests in software engineering for RDF data by checking an RDF graph against test cases defined with SPARQL.
Another direction is to using constraints directly as part of the schema language. OWL Flight \cite{DBLP:conf/www/BruijnLPF05} has been a proposed OWL dialect that expresses cardinality and range constraints on properties. Integrity Constraints in OWL \cite{DBLP:conf/aaai/TaoSBM10} describes an integrity constraint semantics for OWL restrictions to enable validation. 
As a result of the standardization efforts for RDF validation, SHACL \cite{bibSHACL} has been recommended by the W3C. SHACL facilitates the creation of data shapes for validating RDF data. SHACL specification does not define a formal semantics, even though there are some proposals \cite{BonevaS2016, cormanshacl2018}. SHACL shapes are informally converted into SPARQL queries. Similarly, the Shape Expressions (ShEx) \cite{prud2014shape, DBLP:conf/semweb/BonevaGP17} language is another shape language with an abstract syntax and formal semantics. 

 Domain specific patterns can also be seen from the perspective of Ontology Design Patterns (ODPs).  The main goal of ODPs is to improve the process of ontology development and its re-usability, linking data publishing, knowledge extraction, and knowledge engineering  \cite{Gangemi_ODBforSWContent2005, hitzler2016ontology}. 
Content ODPs are used to solve recurrent content modelling problems \cite{hammar2017content}. The ontology design patterns  can also be used to generate SHACL shapes for data validation over RDF datasets \cite{panditusing2018}. 

The domain specification approach proposed in this paper aims to bring a compact, formal solution for customizing a very large and heterogeneous vocabulary, the de-facto content annotation standard schema.org, for specific domains and tasks. Although such endeavor is related to RDF validation, the main focus here is not to validate a graph but rather create machine readable patterns for content and data annotations. The investigation of how the proposed approach can be used for creating Content ODPs is left as future work. We adopt the SHACL syntax in order to benefit its uptake and increasing tool support.



\section{Conclusion and Future Work}
\label{sec:conclusion}
Schema.org is the de-facto industrial standard for annotating content and data. Due to its design, it is not very straight forward for data publishers to pick right types and properties for specific domains and tasks. To overcome this challenge, machine-readable domain specific patterns can guide the knowledge generation process and in a larger scale, the curation of a knowledge graph.

In this paper, we proposed a formal approach, the domain specification process for generating aforementioned domain specific patterns. The domain specification process applies an operator implemented in SHACL to schema.org vocabulary, in order to reduce its generic complexity and restrict it for certain tasks. We presented our approach by formalizing different types of domain specification processes and giving an abstract syntax for domain specification operators based on SHACL. We demonstrated the utility of our approach by various use cases and examples.

For the future work, we will work on specifying the fourth domain specification process, CDS, for defining more complex constraints such as inter-property value relations. Additionally, we will introduce frame-based semantics for the validation task. We will also improve the tooling for both creating domain specification operators and using them in generation and curation tasks.

%
%

 \bibliographystyle{splncs04}
 \bibliography{references}

\end{document}